\documentstyle[preprint,aps,epsf,eqsecnum]{revtex}

\begin{document}
\tighten

\newcommand{\be}{\begin{equation}} \newcommand{\ee}{\end{equation}}
\newcommand{\bea}{\begin{eqnarray}} \newcommand{\eea}{\end{eqnarray}}
\newcommand{\r}{\rangle } \newcommand{\la}{\langle }

\def\Re{{\rm Re}}
\def\Im{{\rm Im}}
\def\PiL{\Pi_{\rm L}}
\def\PiT{\Pi_{\rm T}}
\def\rhoT{\rho_{\rm T}}
\def\rhoL{\rho_{\rm L}}
\def\PL{P_{\rm L}}
\def\PT{P_{\rm T}}

\def\j{{\bf j}}
\def\x{{\bf x}}
\def\p{{\bf p}}
\def\q{{\bf q}}
\def\k{{\bf k}}
\def\v{{\bf v}}
\def\u{{\bf u}}
\def\E{{\bf E}}
\def\D{{\bf D}}
\def\magp{|{\bf p}|}
\def\f{{\overline{f}}}

\def\A{{\rm A}}
\def\B{{\rm B}}
\def\V{{\rm V}}
\def\T{{\rm T}}
\def\LHS{{\rm LHS}}
\def\naBla{{\bf \nabla}}
\def\ij{{i \cdots j}}

\draft



\title {Shear viscosity of hot  QCD from transport theory and thermal field theory in real time formalism}

\author{  Hou Defu}

 \address{  Institute of Particle Physics, Huazhong Normal University, 430079 Wuhan,
China \\
 Department of Physics, The Ohio state University,  Columbus, OH 43210,USA\\
Institut f\"ur Theoretische Physik,
J.W. Goethe-Universit\"at, D-60054 Frankfurt/Main, Germany}

 \date{\today}
\maketitle

\begin {abstract}

We study shear viscosity in weakly coupled hot pure gauge field QCD theory basing on
transport theory and the Kubo formula using the closed time path formalism (CTP) of real time finite
temperature field theory. We show that the viscosity can be obtained as the integral
of a  retarded three-point function. Non-perturbative corrections to the bare one
loop result  can be obtained by solving a Schwinger-Dyson type integral equation for
this vertex. This integral equation represents the resummation of an infinite 
series of ladder diagrams which all contribute to the leading-log order result. We
show that this integral equation has exactly the same form as  the
linearized Boltzmann equation and explain the reason behind this formal equality.

\end{abstract}


\narrowtext

\section{Introduction}

    Fluctuations occur in a system slight perturbed away from equilibrium. The responses to these fluctuations are described by transport coefficients which
 characterize the dynamics of long wavelength, low frequency fluctuations in the medium\cite{eh,Groot}.
The investigation of transport coefficients in high temperature
gauge theories is important in cosmological applications
such as electroweak baryogenesis\cite{bg}, and in the context of heavy ion collisions \cite{hvi,az}.

 There are two basic methods to calculate transport coefficients: transport theory and  response theory\cite{gyu,kubo,MartinK,Hosoya,Horsley,hu,llog}.  Using the transport theory method one starts from a local equilbrium form for the distribution function and performs an expansion in the gradient of the four velocity field. The coefficients of this expansion are determined from the classical Boltzman equation
\cite{llog,jeon1}.  
In the response theory approach one divides the Hamiltonian into a bare piece and 
a perturbative piece that is linear in the gradient of the velocity field. 
 One uses standard perturbation theory to obtain the Kubo formula for the 
 viscosity in terms of retarded Green functions\cite{kubo,jeon1}.
   These Green functions are then evaluated using equilbrium quantum field theory. 
    As is typical in finite temperature field theory, it is not sufficient to 
    calculate perturbatively in the coupling constant even for the weak coupling case. 
     There are certain infinite sets of diagrams that contribute at the same order in perturbation theory and have to 
be resummed \cite{jeon1,pisarski,meg}.

In this paper, we want to compare these two methods.  The response theory approach
allows us to calculate transport coefficients from first principles using the well
understood methods of quantum field theory. On the other hand, the transport theory
approach involves the use of the Boltzman equation which is itself derived from some
more fundamental theory using, among other things, the quasiparticle approximation. 
In this sense, the response theory approach is more fundamental than the transport
theory method. However, the response theory approach can be quite difficult to
implement, even for a high temperature weakly coupled scalar theory, because of the
need to resum infinite sets of diagrams\cite{jeon1,pisarski}.  These considerations
motivate us to understand more precisely the connection between the more practical
transport theory method, and the more fundamental response theory approach.  

Some progress has already been made in this direction.  It has been shown that
keeping only terms which are linear in the gradient expansion in the transport
theory calculation is equivalent to using the linear response approximation to
obtain the usual Kubo formula for the shear viscosity in terms of a retarded
two-point function, and calculating that two-point function using standard
equilibrium quantum field theory techniques to resum an infinite set of ladder 
diagrams for scalar fields \cite{jeon1,meg,hz2,meg1}.  This result is not surprising since it
has been known for some time that ladder diagrams give large contributions to
$n$-point functions with ultra soft external lines \cite{bdc,ASY,lm,Blaizot,blz}.   Some
studies on transport coefficients and  ladder resummation for  hot gauge theory  using Kubo formule 
within  the imaginary time formalism  were reported in  \cite{Martinez,Basagoiti,Aarts}. 
Due to the need for resummation, the diagrammatic approach to transport
coefficients in hot field theories, based on Kubo formulae, has 
developed a reputation for being extremely  cumbersome.
This is particularly true for the ITF formalism with its need for
analytical continuation and the corresponding complicated cutting rules \cite{Basagoiti}
As a result, an alternative approach based on the 
Boltzmann equation in kinetic theory has recently become more popular 
\cite{llog}.With this appraoch, the transport cofficients  in high temperature theory has recently been studied up to
leading order in  \cite{llog}. But these calculations are not trivial 
either, and they cannot fully replace the kind of intuitive insight into 
the underlying physical mechanisms which is provided by a diagrammatic analysis. 
 Our goal is therefore to calculate the shear viscosity of hot gauge theory 
from both transport method and diagrammatic method in real time formalism , and thus understand 
the connection of these two methods.

This paper will be organized as follows. In section II we define shear viscosity 
 using a hydrodynamic expansion of the
energy-momentum tensor. In section III we calculate shear viscosity using the transport theory method
by performing a gradient expansion on the Boltzmann equation and obtaining a
linearized Boltzmann equation.   In section IV we  calculate the shear viscosity 
using the  Kubo formula which relates the shear viscosity  to the
two-point retarded Green function of the viscous-shear stress tensor. 
Starting from the Kubo formula we calculate the shear viscosity at leading-log approximation
using standard techniques of finite temperature quantum field theory in real time
formalism. We show that the shear viscosiy can be obtained as an integral over a
three-point vertex. This three-point vertex  satisfies an integral equation,
 which  represents the resummation
of an infinite series of ladder diagrams. We show that this
integral equation representating ladder resummations  has the
same form as the linearized Boltzmann equation and explain the reason behind this formal equality.  
 We discuss our results and present our conclusions section V.

\section{Viscosity}

In a system that is out of equilibrium, the existence of gradients
in thermodynamic parameters like the temperature and the four dimensional velocity field give rise to thermodynamic forces.   These thermodynamic forces lead to  deviations from the equilibrium expectation value of the energy momentum tensor which are characterized by transport coefficients like the thermal conductivity and the shear and bulk viscosities.   In order to separate these different physical processes we decompose the energy-momentum tensor as,
\begin  {equation}
T^{\mu\nu}=\epsilon u^\mu u^\nu-p\Delta^{\mu\nu}+P^\mu u^\nu +P^\nu u^\mu +\pi^{\mu\nu}\,;~~~~\Delta_{\mu\nu}=g_{\mu\nu}-u_\mu u_\nu.
\end{equation}
The quantities $\epsilon$, $p$, $P_\mu$ and $\pi_{\mu\nu}$ have the physical meanings of internal energy density, pressure, heat current and viscous shear stress, respectively.  The four vector $u_\mu(x)$ is the four dimensional velocity field which satisfies $u^\mu(x) u_\mu(x)=1$.  The expansion coefficients are given by
\begin {eqnarray}
\epsilon&=&u_\alpha u_\beta T^{\alpha\beta},\nonumber\\
p&=&-\frac{1}{3}\Delta_{\alpha\beta}T^{\alpha\beta},\nonumber\\
P_\mu&=&\Delta _{\mu\alpha}u_{\beta} T^{\alpha\beta}, \label{def1}\\
\pi_{\mu\nu}&=&(\Delta_{\mu\alpha}\Delta_{\nu\beta}-\frac{1}{3}\Delta_{\nu\mu}
\Delta_{\alpha\beta})T^{\alpha\beta}.\nonumber
\end{eqnarray}
The viscosity is obtained from the expectation value of the viscous shear stress part of the energy momentum tensor.  We expand in gradients of the velocity field and write, 
\begin{eqnarray}
&& \delta \langle  \pi_{\mu\nu}\rangle =\eta H_{\mu\nu} +  \cdots   \nonumber \\
&& H_{\mu\nu} = \partial_\mu u_\nu + \partial_\nu u_\nu - \frac{2}{3} \Delta_{\mu\nu} \Delta_{\rho\sigma}
\partial^\rho  u^\sigma 
\label{DEF}
\end{eqnarray}
where $\eta$  is the coefficient of the term that is linear  in the gradient of the four velocity, defined as shear viscosity.

Throughout this paper we work with weak-coupled  $SU(3)$ Yang-Mills theory. 
 The Lagrangian for this theory is
\begin {eqnarray}
{\cal L} &=& \frac{1}{4} F_{\mu\nu}^a F^{\mu\nu}_{ a}\nonumber
\\
F_{\mu\nu}^a &=&\partial_\mu A_\nu^a - \partial_\nu A_\mu^a - ig f^{abc}A_\mu^b
A_\nu^c
 \end{eqnarray}
where the coupling constant $g\ll 1$. In local rest frame, the traceless spatial part of the energy-momentum tensor is

\begin {eqnarray} \pi_{ij} =  F_{i \mu}^a F^a _{\mu j}- \frac{1}{3} \delta_{ij} 
F^a_{\alpha\mu} F^a_{\alpha\mu}   \end{eqnarray}


\section {Shear viscosity from kinetic theory }

   
    Kinetic  theory and the Boltzman equation can be used to study transport
properties of dilute many-body systems.  One assumes that, except during brief 
collisions, the system can be considered as being composed of classical particles
with well defined position, energy and momentum.  This picture is valid when the
mean free path is large compared with the Compton wavelength of the particles.  At
high temperature, the typical mean free path of thermal excitations in hot QCD is ${\cal
O}(1/g^2T \ln g^{-1})$ and is always larger than the typical Compton wavelength of
effective thermal oscillations which is ${\cal O}(1/{gT})$ . We introduce a phase
space distribution function $f(x,\underline{k})$ which describes the evolution of
the phase space probability density for the fundamental particles comprising a
fluid. In this expression and in the following equations the underlined momenta  are
on shell.

The Boltzmann equation  of  pure  gauge field QCD describing the evolution of the
gluon distribution function $f(x,k)$  has the form:

\begin {equation}
    \left[
	{\partial \over \partial t}
	+
	\v_\k \cdot {\partial \over \partial \x}
	+
	{\bf F_{\rm ext}} \cdot {\partial \over \partial \k}
    \right]
    f(\k,\x,t)
    =
    C[f] \,.
\label {eq:Boltz}
\end{equation}   
The external force ${\bf F}_{\rm ext}$ term will only be relevant
in discussing the electrical conductivity . For calculations to leading logarithmic order in $g$,
 and for the transport
coefficient under consideration, it will be sufficient to
include in the collision term $C[f]$ only two-body scattering processes,
so that

\begin {eqnarray}
{\cal C}[f] = \frac{1}{2} \int_{123} d \,\Gamma_{12\leftrightarrow 3k}[
f_1 f_2 (1+f_3)(1+f_k) - (1+f_1)(1+f_2) f_3 f_k ] \label{BTZE}
\end{eqnarray}
with $f_i:= f(x,\underline{p}_i)$, $f_k:=f(x,\underline{k})$.  The symbol $d \, \Gamma
_{12\leftrightarrow 3k}$ represents the differential transition rate for particles of
momentum $P_1$ and $P_2$ to scatter into momenta $P_3$ and $K$ and is given by
\begin {eqnarray}
d\,\Gamma_{12\leftrightarrow 3k} := \frac{1}{2\omega_k} |{\cal
T}(\underline{k},\underline{p}_3,\underline{p}_2,\underline{p}_1|^2
\Pi^3_{i=1} \frac{d^3  p_i}{(2\pi)^32\omega_{p_i}} (2\pi)^4
\delta(\underline{p}_1 + \underline{p}_2 - \underline{p}_3 - \underline{k})
\end{eqnarray}
where ${\cal T}$ is the multiparticle scattering amplitude.

The form for
$f(x,\underline{k})$ in local equilibrium is, 
\begin {equation}
f^{(0)}=\frac{1}{e^{\beta(x) u_\mu(x) \underline{k}^\mu}-1}:=n_k,;~~N_k:=1+2n_k
\label{fequib} 
\end{equation}  

We study the Boltzmann equation in the hydrodynamic
regime where we consider times which are long compared to the mean free time and
describe the relaxation of the system in terms of long wavelength fluctuations in
locally conserved quantities.  
For a simple fluid without any additional conserved
charges, the only locally conserved quantities are energy and momentum.   To solve
the Boltzman equation in this near equilibrium hydrodynamic regime, we expand the
distribution function around the local equilibrium form using a gradient expansion. 
We go to a local rest frame in which we can write $\vec{u}(x)=0$.  Note that this
does not imply that gradients of the form $\partial_i u_j$ must be zero.  In the
local rest frame (\ref{DEF}) becomes, 

\begin {eqnarray} && \delta \langle 
\pi_{ij}\rangle = -\eta  H_{ij}  + \cdots  \label{DEFB} \\ && H_{ij} = \partial_i
u_j + \partial_j u_i - \frac{2}{3} \delta_{ij} (\vec{\partial} \cdot \vec{u})
\nonumber \nonumber \end{eqnarray} In all of the following expressions we keep only
linear terms that contain one power of $H_{ij}$ and quadratic terms 
that contain two
powers of $H_{ij}$, since these are the only terms that contribute to the shear 
viscosity we are trying to calculate.  

We write,
\begin {equation}
f=f^{(0)} + f^{(1)}  + \cdots 
\label{expf}
\end{equation}   
with,
$f^{(1)} \sim \underline{k}_\mu \partial^\mu f^{(0)}$.
Using (\ref{fequib}) we obtain,
\begin {eqnarray}
&&\underline{k}_\mu \partial^\mu f^{(0)} = \beta\, n_k(1+n_k)\,\frac{1}{2} I_{ij}(k) H_{ij} \nonumber \\
&& f^{(1)} := - n_k(1+n_k) \phi_k\,;~~ \phi_k =\beta \frac{1}{2}B_{ij}(\underline{k})H_{ij}  \label{def2}
\end{eqnarray}
where we define
\begin {eqnarray}
\label{def3}
&& \hat I_{ij}(k) = 
(\hat k_i \hat k_j-\frac{1}{3}\delta_{ij})\,;~~I_{ij}(k) = k^2\hat I_{ij}(k)
\end{eqnarray} 
and write,
\begin {equation}
B(\underline{k})_{ij} = \hat I_{ij}(k) B(\underline{k})
 \label{def4} 
\end{equation}   

The viscous shear stress part of the energy momentum tensor is given by 
\begin {equation} 
\label{defv}
\langle  \pi_{ij}\rangle  = \int \frac{d^3 k}{(2\pi)^3 2\omega_k}  f(x,\underline{k}) \, (k_i k_j-\frac{1}{3}\delta_{ij} k^2)
\end{equation}   
Using the expansion (\ref{expf}) and  rotation invariance
\begin {eqnarray}
&& k_i k_j B(\underline{k}) ~~ \rightarrow ~~ \frac{1}{3} \delta_{ij} k^2 B(\underline{k}) \nonumber \\
&& k_i k_j \hat k_l \hat k_m B(\underline{k}) ~~\rightarrow~~ \frac{1}{15}(\delta_{ij}\delta_{lm}+\delta_{il}\delta_{jm}+
\delta_{im}\delta_{jl}) k^2 B(\underline{k}) 
\end{eqnarray}
 we obtain the linear  contribution from (\ref{def2})  
\begin {eqnarray}
&& \delta\langle \pi_{ij}\rangle =  -\frac{\beta}{15} \int \frac{d^3 k}{(2\pi)^3 2\omega_k} n(1+n) k^2B(\underline{k})\,H_{ij}.   
\end{eqnarray}
The lowest order term gives zero. Comparing with (\ref{DEFB}) we have,
\begin{eqnarray}
\eta = \frac{\beta}{15} \int \frac{d^3 k}{(2\pi)^3 2\omega_k} n(1+n) k^2B(\underline{k}) \label{ttf}
\end{eqnarray}
Thus we have shown that the shear viscosity  can be obtained from the function  $B(\underline{k})$ .  This
 function are the coefficient of the linear  term in the gradient expansion of
  the distribution function.  Next  we will show that this  functions can be
   obtained from the  linearized  Boltzmann equation.


Using the expansion (\ref{expf}) and (\ref{def2}) one can linearize  Boltzmann equation
\begin {equation}
\underline{k}^\mu\partial_\mu f^0 (x,\underline{k})={\cal C}[f^{(0)};f^{(1)}]
\end{equation}   
where we keep terms linear in $f^{(1)}$ on the right hand side.
The left hand side gives,
 
\begin {eqnarray}
\underline{k}^\mu\partial_\mu f^0 (x,\underline{k})
= - n_k(1+n_k)\beta\,\frac{1}{2} I_{ij}(k) H_{ij}
\end{eqnarray}
The right hand side is 
\begin {equation}
{\cal C}[f^{(1)}]= -\frac{1}{2} \int_{123} d \,\Gamma_{12\leftrightarrow 3k} (1+n_1)(1+n_2)n_3 n_k(\phi_k+\phi_3-\phi_1-\phi_2)
\end{equation}   
Using the definition of $\phi$ given in (\ref{def2}) and comparing the coefficients of $H_{ij}$ on both side of the Boltzman equation we obtain,

\begin {eqnarray}
I_{ij}(k)=\frac{1}{2} \int_{123} d \,\Gamma_{12\leftrightarrow 3k}
\frac{(1+n_1)(1+n_2)n_3}{1+ n_k}[   B_{ij}(\underline{p}_3) 
+ B_{ij}(\underline{k})  -B_{ij}( \underline{p}_1)
 - B_{ij}(\underline{p}_2) \,]
\label{Bint}
\end{eqnarray}
This result is an inhomogeneous linear integral equation which can be solved self-consistently to obtain the function $B_{ij}(\underline{k})$.

\section{Viscosity and ladder resummation  in real time formalsim  }

 
The Kubo formulae allow us to use quantum field theory to calculate nonequilibrium transport coefficients.  The results should be the same as those obtained in the previous section using transport theory.  
The Kubo formula  expresses the shear viscosity in terms of a retarded
 two-point Green function \cite{kubo,jeon1,meg}

\begin {equation}
\eta = \frac{1}{10}\frac{d}{d q_0}{\rm Im} [ \lim_{\vec{q} \to 0}G_R(Q)]|_{q_0=0} 
\label{kb01}
\end{equation}
where $G_R$ is the retarded  two-point Green function of the viscous shear stress

\begin {equation}
G_R(x,t;x',t'')=-i\theta(t-t'')[\pi_{ij}(x,t),\pi_{ij}(x',t'')]
\end{equation}   

We obtain a diagramatic expansion for the viscosity from (\ref{kb01})
 .    We use the closed time path formulation of finite temperature field theory, and work in the Keldysh representation.  Several reviews of this technique are available in the literature \cite{keld,schw,Chou,rep-145,PeterH}.  The closed time path integration contour involves two branches, one running from minus infinity to 
positive infinity just above the real axis, and one running back from positive infinity to 
negative infinity just below the real axis\cite{Chou,PeterH}.  All fields can take values on either branch of 
the contour and thus there is a doubling of the number of degrees of freedom. It is straightforward to show that this doubling 
of degrees of freedom is necessary to obtain finite Green functions.

The two-point function or the propagator can be written as a $2 \times 2 $
matrix of the form
 \begin{equation}
 \label{2}
   D = \left(  \matrix {D_{11} & D_{12} \cr
                        D_{21} & D_{22} \cr} \right) \, ,
 \end{equation}
where $D_{11}$ is the propagator for fields moving along $C_1$,
$D_{12}$ is the propagator for fields moving from $C_1$ to $C_2$, etc.

The retarded and advanced propagators are given by the combinations,
\bea \label{3a}
   D_R &=& D_{11} - D_{12}  \nonumber\\
   D_A &=& D_{11} - D_{21} \, .
\eea
The 1PI part of the two-point function, or the polarization insertion, is
obtained by truncating legs.  The retarded and advanced parts are given by,
\bea
 \Pi_R &=& \Pi_{11} + \Pi_{12}  \nonumber\\
 \Pi_A &=& \Pi_{11} + \Pi_{21} \, .
\eea

The situation is similar for higher $n$-point functions.  For example, the
three-point function which is retarded with respect to the first leg is given
by
\bea
\Gamma_{R1} = \Gamma_{111} +\Gamma_{112} +\Gamma_{121} +\Gamma_{122}\,.
\eea
The other three-point vertices that we will  need are:
\bea
&& \Gamma_{R2} = \Gamma_{111} +\Gamma_{112} +\Gamma_{211} +\Gamma_{212}
\nonumber \\
&& \Gamma_{R3} = \Gamma_{111} +\Gamma_{121} +\Gamma_{211} +\Gamma_{221}
\nonumber \\
&& \Gamma_{F} = \Gamma_{111} +\Gamma_{121} +\Gamma_{212} +\Gamma_{222}
\nonumber 
\eea

In our case, we want to obtain a perturbative expansion for the correlation
 functions of composite operators   $D_R(X,Y)$  which appear in (\ref{kb01}).  We define the vertices $\Gamma_{ij}$  as the vertice obtained by
 truncating external legs from the following connected vertice :

\begin {eqnarray}
 \Gamma^{C(ab)}_{ij,\mu\nu} = \langle T_c \pi_{ij}(X)  A^a_\mu(Y) A^b_\nu(Z)\rangle  
\end{eqnarray}
where $T_c$ is the operator that time orders along the closed time path contour.
From this definition, one can draw the skeleton diagram of  the two-point correlation 
function  of the  the viscous shear stress tensors with a  full vertice and full propagators as

\begin {eqnarray}
\parbox{14cm}
{{
\begin {center}
\parbox{10cm}
{
\epsfxsize=8cm
\epsfysize=5cm
\epsfbox{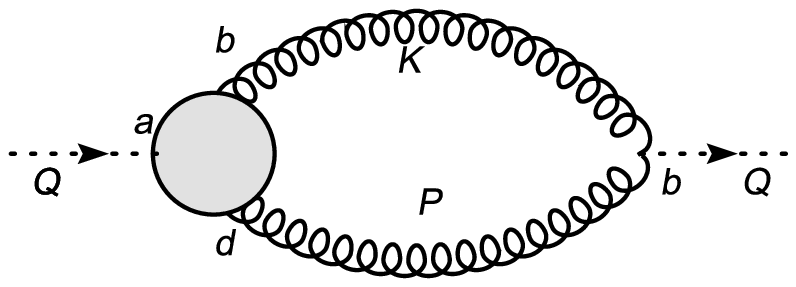}}\\
\parbox{14cm}{\small \center  Fig.~1:  Skelon diagram of the two-point correlation function
 for shear viscosity from linear response. The dashed external line represents the composite
  operator $\pi_{ij}$  }
\label{F01}
\end{center}
}}
\nonumber
\end{eqnarray}

The coupling vertex between the operator $\pi^{ij}$ and two gluons with
incoming
momenta $P,K$ and indices $(\mu,a),(\nu,b)$ respectively can be read
\begin {eqnarray}
 &&\Gamma_{\mu\nu,ij}^{ab(0)}(P,K)=-\delta^{ab}[\delta^{\mu\nu}
\left(
 p^{i}k^{j}+k^{i}p^{j}-\frac{2}{3}\delta^{ij}\p \cdot \k
\right)
+P\cdot K\left(
\delta^{i\mu}\delta^{j\nu}+\delta^{i\nu}\delta^{j\mu}-\frac{2}{3}\delta^{ij}\delta^{r\mu}\delta^{r\nu}\right)
  \nonumber \\ 
&&-\left(
p^{i}K^{\mu}\delta^{j\nu}+p^{j}K^{\mu}\delta^{i\nu}+P^{\nu}k^{i}\delta^{j\mu}
+P^{\nu}k^{j}\delta^{i\mu}
\right)
-\frac{2}{3}\delta^{ij}\left(P^{\nu}k^{r}\delta^{r\mu}+K^{\mu}p^{r}\delta^{r\nu}\right)],
\end{eqnarray}
where $P=(p_{0},\p)$, $P\cdot K=p_{0} k_{0}-\p \cdot \k$ and
$p_{0}$
is the energy of the corresponding gluon.

For $Q=0$,$P=-K$ and on shell gluons ,$K^2=0$, $\Gamma(K,Q,P)\equiv \Gamma(K)$,
 we have:

\begin {eqnarray}
P_{\mu\nu}^T (K) \Gamma^{ab (0)}_{ij,\mu\nu} (K)&=&-2 \delta^{ab} I_{ij}(k),
\nonumber
\\
P_{\mu\alpha}^T(K) \Gamma^{ab (0)}_{ij,\alpha\beta} (K)P_{\nu\beta}^T(K) 
 &=&-2 \delta^{ab} I_{ij}(k)P_{\mu\nu}^T(K)
\label{prj1}
\end{eqnarray}
where $P_{\mu\nu}^T(K)$ is the transverse projector defined as $P_{00}^T=P_{0i}^T=0$, and
$P_{ij}^T(k)=\delta_{ij}-\hat \k_i\hat \k_j$. $\hat \k_i$ represents unit vector of $\k$.
Rotation invariance leads for the full vertice:

\begin {eqnarray}
 P_{\mu\nu}^T (K) \Gamma^{ab }_{ij,\mu\nu} (K)&=&-2 \delta^{ab}
\Gamma(K) \hat I_{ij}(k) ,\nonumber \\ P_{\mu\alpha}^T(K) \Gamma^{ab }_{ij,\alpha\beta}
(K)P_{\nu\beta}^T(K)  &=&-2 \delta^{ab} \hat I_{ij}(k)P_{\mu\nu}^T(K)\Gamma(K)
\label{prj2} \end{eqnarray}

Since for the hard gluons, only the transverse part dominates the leading-log
contribution, we can replace the full gluon propagtor by its transverse part.

\begin {equation}
D^{\alpha\mu}(K)\rightarrow  P^{\alpha\mu}_T D(K)
\end{equation}   
with
\begin {equation}
D(K)=\frac{1}{K^2-\Pi_T(K)}.
\end{equation}   
where $\Pi_T(K)$ represents the transverse gluon self-energy.

Using  the Feynman rules of the closed time path formalism (CTP) of real time finite temperature field theory,we write the two-point correlation function of
the composite operators $\pi_{ij}$
  as

\begin {equation}
 G_{ab}(Q) = i \int dK  \Gamma^{ij,\mu\nu}_{cad}(K,Q,-K-Q) iD^{\alpha\mu}_{bc}(K) 
 iD^{\nu\beta}_{db}(K+Q) \Gamma^{(0){ij,\alpha\beta}}(p) \tau_b 
\end{equation}
shown diagramatically in Fig.1. Where $\int dK\equiv \int \frac{d^4k}{(2\pi)^4} $ , and 
   the indices $\{a,b,c,d\}$ are Keldysh indices and 
take values $\{1,2\}$. The gloun propagators here are full resummed propagators.  We perform the sum over Keldysh indices using the Mathematica program described in \cite{johnS}.

After performing the sums over Keldysh indices and  making use of Eqs (\ref{prj1}),(\ref{prj2}), we find imaginary part of the retarded two-point function

\begin {eqnarray}
{\rm Im} G_R(Q) =-\frac{4}{3} \int dK k^2 (N_{k+q} - N_k)
 [\Gamma_{R2}(K,Q,-K-Q) D_A(K) D_R(K+Q) +h.c ] \label{kubo1} 
\end{eqnarray}

In obtaining this result we have used the fact that we will eventually take the
limit $Q\rightarrow 0$ to obtain the viscosity~(\ref{kb01}). This limit gives
rise to what is known as the pinch effect:  using an obvious notation we  write the pairs of
 propagators $D_A(K) D_R(K+Q)$ 
 as $a_ k r_{k+q}$, terms with a product of factors
$a_k r_{k+q}$ and $r_k a_{k+q}$ contain an extra factor in the denominator that is proportional to the
imaginary part of the self-energy relative to terms with products of propagators $a_k a_{k+q}$ or $r_k r_{k+q}$. Thus terms
proportional to $a_k a_{k+q}$ or $r_k r_{k+q}$ can be dropped. This is the so-called pinch limit, because 
the large terms occur when the contour is ``pinched'' between the poles of the two propagators,
which gives rise to this enhencement factor in the denominator that is proportional to the
imaginary part of the self-enegy.  We regulate the pinching singularity with the imaginary part of the self-energy and obtain \cite{meg},
\begin {eqnarray}
r_k a_{k+q} \rightarrow  -\frac{\rho_k}{2\rm{Im} \Pi^T_k}\,;~~~~\rho_k = i(r_k-a_k) \label{pinch}
\end{eqnarray}
where $\Pi^T$ is the retarded part of the transverse  gluon self-energy.

Now we expand in $q_0$  and keep only the term proportional to $q_0 $ since this term is the only one that
 contibutes to (\ref{kb01}) .  There is  an over all thermal factor of the form $N_k-N_{k+q}$  in
 (\ref{kubo1}).  The expansion of this thermal factor is straightforward:
\begin {eqnarray}
N_k-  N_{k+q} = 2q_0 \beta n_k(1+n_k) + \cdots \label{nexp}
\end{eqnarray} 
Consider the behaviour of the vertices when $q_0\rightarrow 0$.  We introduce the notation
 $\Gamma(K,0,-K):=\Gamma(K)$ .
  It is straightforward to show that 
\begin {eqnarray}
\Gamma_{R2}(K,Q,-K-Q) \rightarrow  {\rm Re}\Gamma_{R2}(K) \label{re}
\end{eqnarray}
Using these results to simplify (\ref{kubo1})  and substituting into (\ref{kb01}) we
obtain, 
\begin{equation} 
\label{vsf}
\eta = \frac{\beta}{15} \int dK\, k^2 \,\rho_k n_k (1+n_k)
\left[\frac {\rm Re \Gamma_{R2}(K)}{\rm Im \Pi^T_k}\right] 
\end{equation}
Comparing
with (\ref{ttf}) we see that the results are identical if we use (\ref{def4})and 
identify

 \begin {eqnarray} 
 \label{compare} 
 &&B(\underline{k}) =
\frac {\rm Re \Gamma_{R2}(\underline{k})}{\rm Im \Pi^T_k}
\label{EQ1}
\end{eqnarray} 
 where momentum $K$ must be on new
mass shell: $\delta(K^2-{\rm Re} \Pi^T_K)$


It has been known for some time that in scalar theory, the set of diagrams which give the dominate
contributions to the viscosity are the ladder diagrams.  These diagrams contribute to the viscosity to 
the same order in perturbation theory as the bare one loop graph, and thus need to be resummed 
\cite{jeon1,jeon2,meg}.  Following the similar power counting rule one can show that the set of
 ladder diagrams shown in Fig.2  give the  dominate contributions to the viscosity in pure gauge
 field QCD theory at high temperature.  This effect
occurs for the following reason. It appears that the ladder graphs are suppressed
relative to the bare vertex  by extra powers of the coupling, which come from the
extra vertex factors that one obtains when one adds rungs (vertical lines). However,
these extra factors of the coupling are compensated for by a kinematical factor. 
This factor arises through the pinch effect which is described above.  The addition
of an additional rung in a ladder graph always produces an extra pair of propagators
of the form $a_k r_{k+q}$ or $r_k a_{k+q}$. Products of this form contribute  a
factor which produces an enhancement.  This factor  occurs when the contour is
``pinched'' between the poles of the two propagators which gives rise to a
contribution in the denominator that is proportional to the imaginary part of the
self-energy.

In order to include those ladder diagrams, which dominate the contribution to the viscosity,
 we obtain the vertex $ \Gamma_{ij}(P,Q,-P-Q)$ as the solution to the integral equation shown in Fig.2.  

\begin {eqnarray}
\parbox{14cm}
{{
\begin {center}
\parbox{10cm}
{
\epsfxsize=10cm
\epsfysize=4cm
\epsfbox{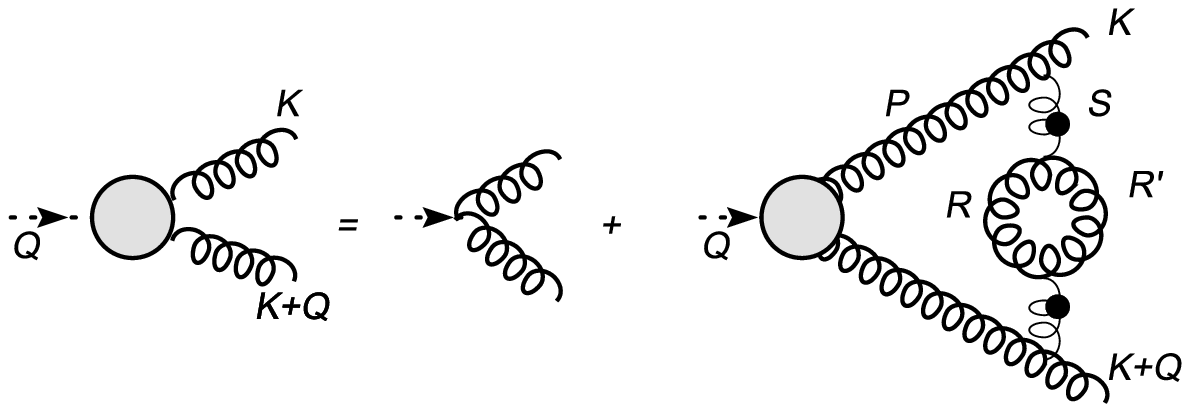}}\\
\parbox{14cm}{\small 
 Fig.2: Integral equation for ladder resummation, the blob stands for the
full vertex  $ \Gamma$ . Thick lines represent for hard transverse gluons, where a finite width is needed to regularize the pinch sigularity. While
thin lines represent soft gluons, where HTLs resummed propagators are required.}
\label{F02}
\end{center}
}}
\nonumber
\end{eqnarray}

Since we only calculate up to the leading-log order, in order to simplify the calculations, 
several points are in order:

1) We consider weak coupled degrees of freedom ($g\ll 1$);

2)Hard degrees of freedom with the momentum of order $T$  for the rails dominate the contribution;

3)Small momentum transfer ($S\ll gT$),$t$ chanel dominates the  contribution. Namely, the momenta on the
rungs (vertical lines) are soft of order $gT$. 

In order to to get rid of color indices and Lorentz indices, We
 contract  $\delta^{ab} P_{\mu\nu}^T(K)$ with
 both sides of above equation and make use of

\bea
&&\delta^{a b}\,\delta^{a a'}\,\delta^{b b'}\, f^{a' c d}\,\delta^{c e}\,f^{e f g}\,
\delta^{f f'}\,\delta^{g g'}\, f^{f' e' g'}\,\delta^{e' c'}\,f^{b'd'c' }\,
\nonumber
\\
&&= f^{aed}f^{efg}f^{aed}f^{efg}=N_c^2
\eea
and
\bea
&&P^T_{\alpha\beta } (K)\,  \gamma_{ \rho,\tau,\alpha }\,  P^T_{\rho \sigma }(P)
\, \gamma_{\sigma,\tau',\beta }
= 8 P_\tau P_{\tau'}\,\nonumber
\\
&&P^T_{\lambda \delta } (R) \,  \gamma_{ \lambda,\kappa,\lambda' }\,  P^T_{\lambda' \delta' }
(R)\,  \gamma_{ \delta,\kappa',\delta' }
= 8 R_\kappa R_{\kappa'}
\eea
where $\gamma_{\alpha,\beta,\gamma}(P,T,K)=(P-T)_\gamma \, \delta_{\alpha\beta}
+(T-K)_\alpha \, \delta_{\beta\gamma}+(K-P)_\beta \, \delta_{\alpha\gamma}$ comes
from the triple-gluon vertex. 

After some algbra calculations, we obtain a matrix integral equation in
Keldysh space

\begin {eqnarray} 
&&  \Gamma^{lm}_{abc}(K,Q, -K-Q)=\Gamma^{(0),lm} 
 -64 N_c^2 \int dP\,dR\, \Gamma^{lm}_{dbe}(P,Q,-P-Q)  D_{ad}(P)
D_{ec}(P+Q) \nonumber \\ 
&&\cdot D_{ac}(R+K-P) \tau_3^c D_{ca}(R)P_{\tau}
R_{\kappa} D^{\tau\kappa} (T)P_{\tau'} R_{\kappa'} D^{\tau'\kappa'}(T) \tau_3^a 
\end{eqnarray}

We perform the sums over the Keldysh indices using the Mathematica program in
\cite{johnS}  and simplify the result by taking $Q$ to zero, keeping only the
pinching terms, and using (\ref{pinch}).   We obtain, 

\begin {eqnarray} 
\label{gammaintsh1}
 \Gamma_{R2}^{lm}(K) = &&k_m k_l
-\frac{1}{3} \delta_{ml} k^2 +\frac{3 N_c^2 }{2}\int dP\,dR\,dR'\,
(2\pi)^4 \delta^4(P+R-R'-K)\nonumber \\ &&
 |{\cal T}|^2 \rho_p \rho_R\rho_{R'}
\frac{\Gamma_{R2}^{lm}(P)}{{\rm Im}\Pi^T_R(P)} (1+n_p)(1+n_{R})n_{R'}/(1+n_k)
\end{eqnarray}
where  ${\cal T}^2\equiv  16 g^4 | P_{\tau} r_S^{\tau\kappa}R_{\kappa}|^2$, 
which is recognized as the matrix element squared for the collision of the hard gluon in $t$ chanel as shown 
in Fig. 3(a). Here
$r_S^{\tau\kappa}$ is the hard thermal loops resummed retarded propagator for the soft transfer momentum
$S$.

\begin {eqnarray}
\parbox{14cm}
{{
\begin {center}
\parbox{12cm}
{
\epsfxsize=10cm
\epsfysize=4cm
\epsfbox{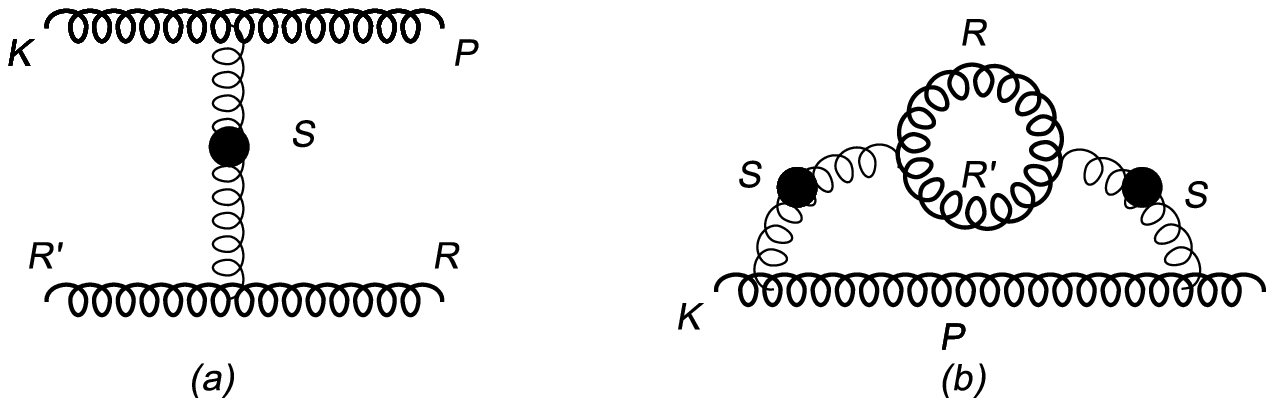}}\\
\parbox{14cm}{\small   Fig.~3:  (a) Two-body  gluon-gluon elastic scattering in $t$ chanel. (b)
Self-energy describing collisions in the (resummed) Born approxximation. The thick lines refer to the
the hard colliding gluons. The thin lines with a blob denote soft gluon propagators 
dressed by the HTL}
\label{F03}
\end{center}
}}
\nonumber
\end{eqnarray}

Note that this integral equation is decoupled: the only three point vertex that
appears is $\Gamma^{lm}_{R2}$. To simplify this expression further we write
$\Gamma_{ij} = \hat I_{ij}(p) \Gamma(P)$, use (\ref{compare}) and symmetrize the
integral on the right hand side over the three integration variables $P$, $R'$ and
$R$. We multiply and divide the left hand side by $\rm{Im} \Pi^T_R(P)$ and replace
this expression in the numerator by the HTL result \cite{Blaizot,dp,thoma}

\begin
{eqnarray} {\rm Im}\Pi^T_R(K) =&&\frac{1}{2}\left(\frac{1}{1+n_k}\right) \int
dP\,dR\ dR'\, (2\pi)^4 \delta^4(P+R-R'-K) |{\cal T}|^2 \nonumber \\
&&\rho_p\rho_r\rho_{R'}(1+n_p)(1+n_{R'})n_R\label{pi} 
\end{eqnarray}

Rearranging we obtain \cite{meg}, \begin {eqnarray} && I(k,k)_{lm} = \frac{N^2}{2}
\int dP\,dR\,dR' ~(2\pi)^4 \delta({P} + {R} - {R'} - {K})|{\cal T}|^2  \frac{
(1+n_p)(1+n_{R'})n_R }{(1+n_k)} \label{Bint2} \\ && ~~~~~~~ \rho_p \rho_r\rho_{R'}
[   B_{lm}({R'}) +B_{lm}({K})-B_{lm}({P}) -B_{lm}({R}) ]  \nonumber \end{eqnarray}

where we have used (\ref{re}) and (\ref{EQ1}).
  When the delta functions are used to
do the frequency integrals, this equation has exactly the same form as the equation
obtained from the  linearized  effective Boltzman equation(\ref{Bint}), with a new
mass shell describing thermal excitations.   Comparing (\ref{ttf}) and (\ref{Bint})
with (\ref{vsf}) and (\ref{Bint2}) we conclude that calculating  shear
viscosity  of the hot  pure gauge field QCD using effective transport theory by keeping only first order terms
in the gradient expansion, is equivalent to using the Kubo formula obtained from
response theory, with a three point vertex obtained by resumming ladder graphs.

Now we have established  an interesting  connection between
   kinetic theory and diagrammatic linear response theory for the nonabilian gauge
   theory. We  obtain formally the same expressions for the viscosity  
   in both approaches.  It is desirable to understand the deeper reasons behind 
   this formal equality. 
   
The viscosity is obtained from the expectation value of the viscous 
shear stress tensor. There
 are two approaches to calculate the expectation value.   The first method is  from  kinetic 
 theory. With this approach, we use a distribution function $f$ to evaluate the the expectation value of
  the viscous shear stress tensor(see Eq.(\ref{defv})). This distribution function is expanded 
 around a local equlibrium function  $f^{(0)}$  up to the first order $f^{(1)}$ of the gradient expansion .
  The  second approach is  by linear response theory. The expectation value  is calculated from
 the non-equilibrium density matrix $\rho\cite{kubo,meg1}$, which can be expanded around the local
 equilibrium density matrix  $\rho_0$  up to the linear response order in gradient expansion , 
 $\rho=\rho_0+\rho_1$. The  expectation value of a physics quantity does not depend on the approaches one uses, either
from the kinetic theory by the distribution function , or from the linear response theory
by the density matrix under the same approximation in gradient expansion.
Therefore  $B(k)$ from (\ref{ttf}) and from (\ref{compare})should be the same although they seem to describe
  completely  two different things! The former one comes from the first-order distribution
  function $f^{(1)}$; The later,  though  related to a retarded 3-point function, is resulted from the
  first-order matrix density $\rho_1$.   Physicaly , they both describe how far the system is 
  perturbated away from the local equilibrium. In fact,  both  methods describe the first-order deviation of the system away from the local
      equilibrium,  one of them  using  the description of the distribution function, while the
      other one using the density matrix. 
  
  On the other hand,  both appraoches also contain the same physics  processes (diagramms), namely two body collisions.
  In order to understand this  point, let's recall that the Boltzmann equation can
  be derived from the Schwinger-Dyson equation of the full 2-point functions in real-time formalism  under certain 
  approximations , like Wigner transformation and truncation schemes etc \cite{Blaizot}.  Different truncation
  schemes (or self-energy insertions ) lead to different collision terms. The Boltzmann
  equation we used only contains $2\to 2$ collision terms, which come from the cut of the  self-energy diagrams
  containing $2\to 2 $ processes as shown in Fig.3 .  
   The cutting rules tell us  that  the cuts of  self-energy diagramms give  physics scattering amplitudes.
   The diagrammatic method from the integral equation of  the full vertex is essentially
   like a resummation for the full 2-point correlation diagram of the stress tensor. Again in this
   resummation, only the  diagramms  containing $2\to 2$  processes are included. 
   
    Because of these reasons above,  we expect  these two approaches should be able to give the same result for any
theories under the same approximation.

\section{ Summary and Conclusions} \label{V}  

We have studied shear viscosity  of the  nonabelian pure gauge theory  using two different methods.  The first method uses standard
transport theory.  We start from a local equilibrium form for the distribution function and perform a gradient expansion.  We
calculate the  shear viscosity by expanding the Boltzman equation and obtaining a linearized Bolzman equation
 that can be solved consistently. The second technique uses  the Kubo formula for linear
response, which allows us to calculate the  shear viscosity   from the  retarded two-point Green
function of the viscous shear stress tensor. The transport theory calculation  involves the use of  the Boltzman equation,
 which is itself obtained from some more fundamental  theory. The response theory calculation uses the well known  methods of equilibrium
finite temperature quantum field theory and is, in this sense, more fundamental.  However, the calculation is complicated by the
need to resum infinite sets of diagrams at finite temperature. 
At leading order, it is well known that a correct calculation of the linear response coefficient in scalar theory involves 
the resummation of ladder graphs.

We have identified precisely which diagrams  in hot pure gauge QCD need to be 
resummed by studying the connection between the transport theory calculation and the response theory 
calculation in real time formalism.
We have shown that calculating  shear viscosity using transport theory by 
keeping terms that are linear in the gradient of the velocity field in the expansion of the Bolzman equation, 
is equivalent to calculating the same shear viscosity  from quantum field theory at finite temperature using 
the linear response Kubo formula with a vertex given by a specific integral equation.
This integral equation  is the same as that recently  found by Arnold, Moore and
Yaffe~\cite{llog}   by analyzing the infrared divergences of the linearized
collision integrals without screened interactions. This integral equation is also the same as the linearized
Bolzmann equation derived by   Blaizot and Iancu  from the Kadanof-Baym equations by using a gradient expansion and 
keeping only linear terms \cite{Blaizot}.  Thus, this fact constitutes a non trivial check of the formalism 
we have used. This integral equation shows that the complete set of diagrams that need to be resummed includes infinite
 ladder graphs.  We explain the reasons behind this formal equality between this two approaches. First, both approaches
  describe the first-order deviation of the system away from the local
      equilibrium,  one of them  using  the description of the distribution function, while the
      other one using the density matrix. Secondly both methods  involve  the same physics  processes  or diagramms
      ( two body collisions).
 Therefore, We expect this interesting  connection between   these two appraoches should exist for any other theories
 under the same approximations.

 There are several directions for future work.  The study here is only valid up 
to leading-log  approximation of the transport coefficients. Going beyond to this approximation,
one needs to include certain inelastic LPM suppressed splitting
 processes and thus needs to resum more complicated diagrams \cite{llog,ld}.
  It would be interesting to generalize  this work to the full leading order of 
   of  the shear viscosity with  real QCD with quark flavors in real time formalism .



\begin {acknowledgments}

We thank Gert Aarts , E. Braaten, M. Carrington, U. Heinz, R. Kobes , D. Rischke, H. C. Ren, and  J.~M.~Mart\'\i nez Resco  
for interesting discussions. We address special thanks to  U. Heinz for useful suggestions and a  critical reading of
 the original manuscript.This work was partly supported by the National Natural Science 
Foundation of China (NSFC) under project Nos. 10005002 and 10135030.
Hou is grateful to the nuclear theory
group in the physics department of Ohio State University for their hospitality
where the main part of this work was completed.   Hou\  acknowledges financial support from 
Alexander von Humboldt-Foundation.  He also appreciates help and support from 
the Institut f\"ur Theoretische Physik of the J.\ W.\ Goethe-Universit\"at.

\end{acknowledgments}


\appendix


\end{document}